\documentclass{appolb}
\usepackage{epsfig}

\begin{document}
\def\pbarp{\mbox{$\rm  \bar{p}p$}}
\def\ptmin{\mbox{$p_{tmin}$}}
\def\sigmatot{$\sigma_{total}$}

\title{Relevance of ultra-soft gluons and $k_t$ resummation for total
cross-sections
\thanks{Presented at FLAVIAnet Workshop, Kazimierz, 23-27 July, 2009}%
}
\author{
A. Achilli and Y.N. Srivastava \vskip -0.5cm
\address{Physics Department and INFN, University of Perugia, Perugia,
Italy}
\vskip 0.2cm R. Godbole \vskip -0.5cm
\address{Centre for High Energy Physics, Indian Institute of Science,
Bangalore, India}
\vskip 0.2cm A. Grau \vskip -0.5cm
\address{Departamento de F\'\i sica Te\'orica y del Cosmos, Universidad
de Granada, Spain} \vskip 0.2cm G. Pancheri \vskip -0.5cm
\address{INFN Frascati National Laboratories, Frascati, Italy}}
\maketitle
\begin{abstract}
Inclusion of down to zero-momentum gluons and their $k_t$ resummation is shown
to quench the too fast rise of the mini jet cross section and thereby
obtain realistic total cross-sections.
\end{abstract}

\section{Introduction}
We  present details of the very high energy behavior of
hadron-hadron and hadron-photon total cross sections in a model
which incorporates soft gluon $k_{t}$-resummation in the infrared
(IR) region. The effects of this resummation are discussed to
highlight the mechanisms responsible for the rate with which 
the total cross section rises with energy.

The Froissart-Martin (FM) bound \cite{froissart,martin} says that
the hadronic cross-sections cannot rise faster than $\ln ^2 s$.
The most common parametrizations of \sigmatot\ for pp processes,
based on unitarity and constraints from analyticity, impose the FM
bound. By contrast, in our QCD based model \cite{GGPS,PLB08,EPJC},
we find an interesting relationship between the IR behavior of
$\alpha_s$ and the rise of \sigmatot.

We first present a short introduction to the soft gluon
$k_t$-resummation technique, then the main features of our eikonal
model based on QCD-minijet cross section and soft gluon
resummation, after which we compare our results with data from
$pp$ and $\gamma p$ processes. Finally, we show how the
$k_t$-resummation affects the asymptotic rise which, while
respecting the FM bound, does not necessarily saturate it.

\section{$k_{t}$-resummation and the IR limit}
To compute the total cross-section one has to integrate over large
values of the impact parameters $b$, which gets linked to the IR
regions of the underlying partonic processes. Here we are
considering (within a semiclassical approach) soft gluon emissions
from the colliding partons and their $k_t$-resummation, whose
Fourier transforms we assume to describe the matter distribution
within the colliding hadrons. The expression for the transverse
momentum distribution resulting from soft gluon radiation emitted
by the colliding quarks is
 \begin{equation}
d^2P({\bf K_\perp})=d^2{\bf  K_\perp} {{1} \over{(2\pi)^2}}\int
d^2 {\bf b}\ e^{-i{\bf K_\perp\cdot b} -h( b,M)} \label{d2p}
 \end{equation}
 with
\begin{equation}
h(b,M) =  \frac{16}{3}\int^M_0
 {{ \alpha_s(k_t^2) }\over{\pi}}{{d
 k_t}\over{k_t}}\ln{{2M}\over{k_t}}[1-J_0(k_tb)]
 \label{hbq}
\end{equation}
where M gives the energy scale of the process, in our case it is
the maximum allowed transverse momentum for a single gluon
emission \cite{greco}. In various QCD problems \cite{ddt,pp} one
splits this expression in two terms
\begin{equation}
\label{h1} h(b,M) = c_0(\mu,b,M)+ \Delta h(b,M),
\end{equation}
 with the first term parameterizing the infrared behavior and
 the second term which is calculable in the perturbative regime
 defined as
\begin{equation}
\label{h} \Delta h(b,M) =
 \frac{16}{3} \int_\mu^M {\alpha_s(k_t^2)\over{\pi}}[1- J_o(bk_t)]
 {{dk_t}\over{k_t}}
  \ln {
  {{2M}\over{k_t}}}.
\end{equation}
Since this integral only runs down to a scale $\mu>\Lambda_{QCD}$,
one uses the asymptotic freedom expression for $\alpha_s$ and one
assumes that  $J_o(bk_t)$ oscillates to zero. In the range
$1/M<b<1/\Lambda$ an effective h is obtained fixing the scale
$\mu=1/b$. Then one finds \cite{last}
 \begin{equation}
e^{-h_{eff}(b,M)} =
 \big{[}
{{ \ln(1/b^2\Lambda^2) }\over{ \ln(M^2/\Lambda^2) }}
\big{]}^{(16/25)\ln(M^2/\Lambda^2)} \label{PP}
\end{equation}

In our model, the integral (\ref{h}) includes zero momentum values
and we investigate their contribution in the very large impact
parameter region. Since in the IR limit, the perturbative
expression for $\alpha_s$\ is not valid, we employ a
phenomenological expression\cite{GGPS}, which is singular but
integrable in the IR limit of the soft gluon integral
\begin{equation}
\label{alphapheno} \alpha_s(k_t^2)={{12 \pi
}\over{(33-2N_f)}}{{p}\over{\ln[1+p({{k_t^2}
\over{\Lambda^2}})^{p}]}}
\end{equation}
with the parameter $1/2<p<1$. For $p=1$ it coincides with
the Richardson potential \cite{Richardson}. In the large-b region
($b>{{1}\over{N_p\Lambda}}>{{1}\over{M}}$), we can obtain an approximate analytic
expression:
\begin{equation}
\begin{centering}
\begin{array}{l}
\label{halphas3} h(b,M,\Lambda)={{8}\over{3\pi}}
\Biggl [ {{{\bar b}}\over{8(1-p)}} (b^2\Lambda^2)^p \left[
2\ln(2Mb)+{{1}\over{1-p}}\right] +\nonumber {{\bar
b}\over{2p}}(b^2\Lambda^2)^p \left[2\ln(Mb)-{{1}\over{p}}\right]\\
\hspace{2.1cm}+{{\bar
b}\over{2pN_p^{2p}}}\left[-2\ln{{M}\over{\Lambda
N_p}}+{{1}\over{p}} \right] + \nonumber {\bar b} \ln
{{M}\over{\Lambda}}\left[\ln {{\ln{{M}\over{\Lambda}}}\over
{\ln{N_p}}}-1+{{\ln{N_p}}\over{\ln{{M}\over{\Lambda}}}} \right]
\Biggr ]
\end{array}
\end{centering}
\end{equation}
where $N_p=(1/p)^{1/2p}>1$ for $p<1$ , and ${\bar b}=12 \pi/(33-2N_f)$.

Our soft gluon resummation approach allows us to obtain a value for
the ``intrinsic'' transverse momentum
\cite{corsetti}, through
\begin{equation}
h_{intrinsic}=  \frac{16}{3}\int^\Lambda_0
 {{ \alpha_s(k_t^2) }\over{\pi}}{{d
 k_t}\over{k_t}}\ln{{2M}\over{k_t}}[1-J_0(k_tb)]\approx
{{b^2}\over{4}}<k_t^2>_{int}
 \label{hbq2}
\end{equation}
with
\begin{equation}
<k_t^2>_{int}={{8}\over{3 \pi}}{{\bar b}\over{(1-p)}}
\left({{1}\over{2(1-p)}}+\ln{{2M}\over{\Lambda}}\right)\Lambda^2,
\end{equation}
which corresponds to  an intrinsic transverse momentum of a few
hundred MeV for $\Lambda$ in the 100 MeV range and $M\le 1$ GeV.

\section{The Bloch-Nordsieck model for total cross-sections }
In our model for the total hadronic cross-section we use the
eikonal formalism which implies multiple scattering and requires
impact parameter distributions of the scattering particles
\cite{GGPS}. In the high energy limit, neglecting the real part of the
eikonal, the total cross-section can be approximated as given by
\begin{equation}
\sigma_{total}= 2 \int d^2{\bf b} [ 1-e^{ -{\cal I} m \chi(b,s) }].
\end{equation}
The imaginary part of the eikonal is related to the average number
of inelastic collisions $n(b,s)$
 \begin{equation}
2{\cal I} m \chi(b,s)={\bar n}(b,s)
\end{equation}
for which we use the expression
\begin{equation}
 {\bar n}(b,s) = n_{NP} (b,s) + n_{hard} (b,s)
\end{equation}
with the non perturbative (NP)  term relevant only for the low-energy description of
$\sigma_{tot}$ and for its normalization, and the hard one
responsible for the high-energy rise. This term is given by
\begin{equation}
 n_{hard} (b,s) = A(b,s)\sigma _{jet} (s)
  \label{nhard}
 \end{equation}
where $\sigma _{jet}$ is the mini-jet cross section describing
high-energy partonic collisions. It drives the rise of the total
cross section
and depends on the parameter $p_{tmin}$, the minimum transverse
momentum of the scattered partons, which separates hard
parton-parton scattering from all other low-$p_t$ processes and on
currently used, DGLAP evoluted, parametrizations for the Partonic
Density Functions.

A(b,s) is the overlap function given by the Fourier transform of the
transverse momentum distribution resulting from initial state  soft gluon
radiation, which breaks the
collinearity of the colliding partons, making the scattering
process less efficient.
\begin{eqnarray}
A_{BN}(b,s)&=&N\int d^2{\bf  K_{\perp}}\  e^{-i{\bf K_\perp\cdot
b}} {{d^2P({\bf K_\perp})}\over{d^2 {\bf K_\perp}}}={{e^{-h( b,q_{max})}}\over{\int d^2{\bf b} \ e^{-h( b,q_{max})}
 }} \nonumber \\ &=&A_0(s) e^{-h( b,q_{max})}
\label{adb}
\end{eqnarray}
The parameter $q_{max}$ is linked to the maximum transverse
momentum allowed by the kinematics for  single gluon emission
\cite{greco}. It should be calculated for every partonic
subprocess but we make the simplifying assumption of assigning it
a value averaged over all the possible subprocesses \cite{GGPS}
and use it in Eq.~(\ref{nhard}) to calculate the rising part of
$\sigma_{total}$.


Fig.1
 shows our results for $proton-proton$ ( band ) compared with other models
 \cite{GGPS,pp_models}. In this case, our central prediction (black line in the band) has been
obtained with  $p_{tmin} = 1.15$ $GeV$, $p = 0.75$ and GRV
densities.  Application of the model to $\gamma  p$ \cite{EPJC} is
shown on the left  panel.

\begin{figure}[htb!]
  \begin{center}
 \includegraphics[height=5.8cm]{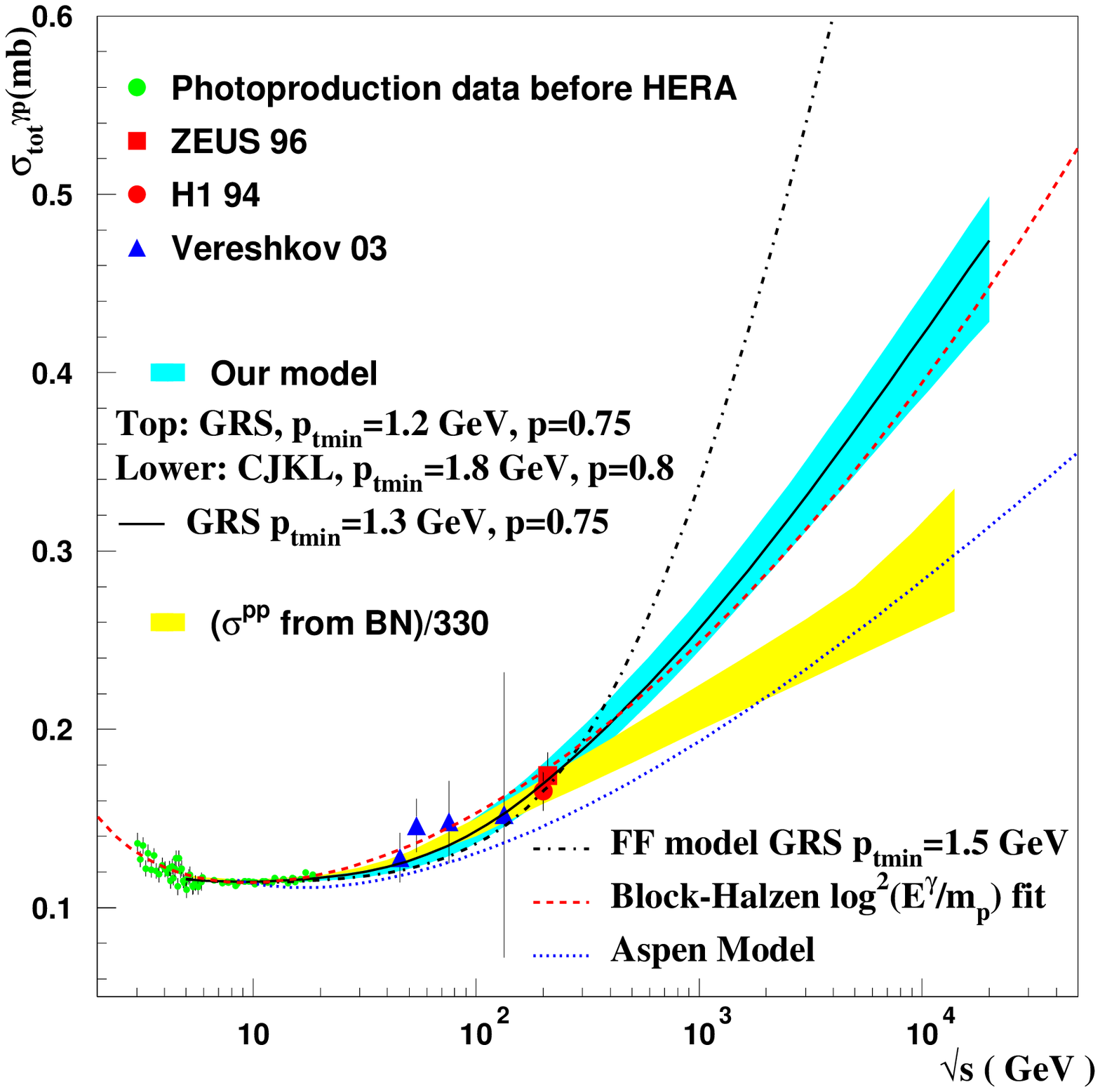}
 \hspace{-0.9cm}
  \includegraphics[height=5.4cm]{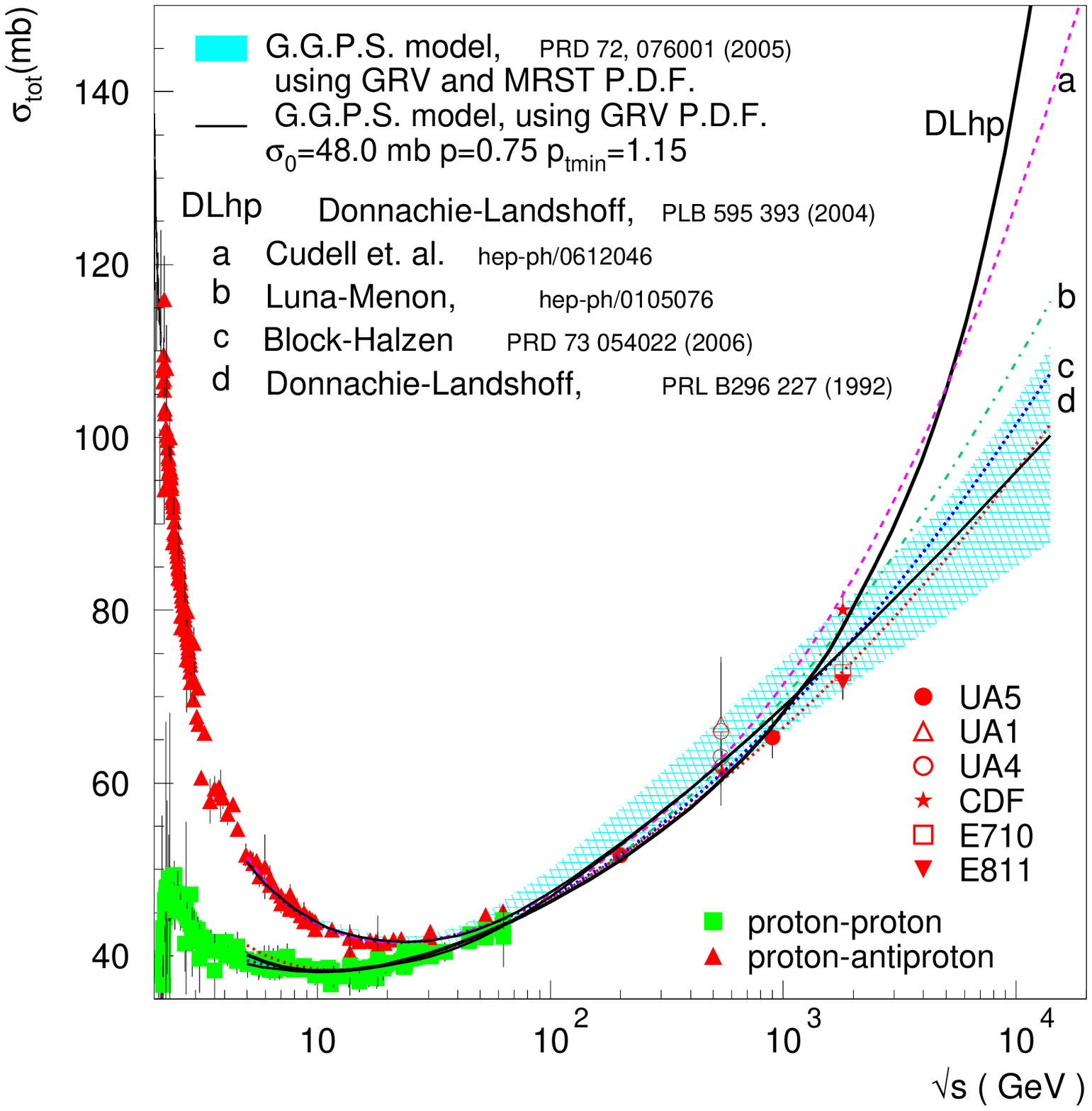}
  \caption{
  $\gamma p$\ total cross section (left) from \cite{EPJC} and
  $pp$\ total cross section (right) from \cite{PLB08} are shown  compared with data and with other
  phenomenological models \cite{pp_models,
  gp_models}.}
  \label{fig:1}
  \end{center}
\end{figure}

\section{Asymptotic limit of the total cross-section}
We can now estimate the very large $s$-limit when the minijet
cross-section rises asymptotically like a power
 $s^\varepsilon$ ($\varepsilon \approx 0.3$ \cite{mpi08}). We find
\begin{equation}
n_{hard} (b,s)=A_{BN}(b,s)\sigma_{jet}(s,p_{tmin})\sim A_0(s)
e^{-h(b,q_{max})}{\sigma_1}({{s}\over{s_0}})^\varepsilon
\end{equation}
From it, we can deduce the very large $b$-limit to be
\begin{equation}
n_{hard} (b,s)\sim A_0(s) \sigma_1 e^{-(b{\bar \Lambda})^{2p}}
({{s}\over{s_0}})^\varepsilon
\end{equation}
with \cite{last}
\begin{equation}
{\bar \Lambda}\equiv
 {\bar \Lambda(b,s)}=\Lambda
 \{
 {{{\bar b}}\over{3\pi(1-p)}}[
 \ln (2q_{max}(s)b) +{{1}\over{1-p}}]
 \}^{1/2p}
 \end{equation}
where $A_0(s)\propto \Lambda^2 $, and $q_{max}$ a very slowly
varying function of $s$. In this limit, the total cross section becomes
\begin{equation}
  \sigma _T (s) \approx 2\pi \int_0^\infty  {db^2 } [1 - e^{ -
C(s)e^{ - (b{\bar \Lambda})^{2p} } } ] \label{sigT}
\end{equation}
with $2C(s) = A_0(s) \sigma _1 (s/s_0 )^\varepsilon$.
We thus obtain
\begin{equation}
\sigma_{T}\approx (constant) [ \ln \frac {s} {s_0}]^{1/p}.
\label{froissart1}
\end{equation}
The constant in Eq.(\ref{froissart1}) depends on $\Lambda$ and $q_{max}$ \cite{last} and is
about one order of magnitude smaller than the constant which
appears in the Martin bound. We thus find that the  introduction of soft gluon
resummation introduces a new scale which imposes a more stringent
 limit on the high energy behavior of \sigmatot.
\section{Conclusions}
In  an eikonal mini-jet model, soft gluon $k_t$-resummation down to
zero gluon momenta does  reduce the strong power-like rise of the
minijet cross-section. We find
 a link between the infrared region in soft $k_t$-resummation
and
the very high energy behaviour of total cross-sections.
Introduction of a dynamical scale related to the very soft gluon
emission,  reduces the
constant occuring in the Martin bound.

\end{document}